\documentstyle[11pt,newpasp,twoside,epsf]{article}
\def\edcomment#1{\iffalse\marginpar{\raggedright\sl#1\/}\else\relax\fi}
\reversemarginpar

\begin{document}
\title{Observations of Galaxies with Chandra}
 \author{G. Fabbiano}
\affil{Harvard-Smithsonian Center for Astrophysics, 60 Garden St.,
Cambridge MA 02138, USA}

\begin{abstract}
This talk discusses the growing impact of {\it Chandra} observations
on the study of galaxies. In particular, we address progress in the 
study of X-ray source populations in diverse galaxies and in the 
study of the hot ISM.
\end{abstract}

\section{Introduction}

This talk is a compendium of early {\it Chandra} (Weisskopf et al 2000)
results, published and in progress,
and a preliminary view of data available in the public {\it Chandra} archive.
As demonstrated by the large amount of discussion on feedback models in this 
meeting, and simulations of {\it Chandra}-like images, the subarcsecond resolution 
of {\it Chandra}, combined with its spectral resolution, is going to provide 
unique answers to the field of galaxy and cluster `ecology' and evolution.
{\it Chandra}'s resolution is optimal for the study of galaxies in X-rays.
{\it Chandra}'s beam is 100 times smaller than that of any past or planned X-ray
mission and  provides a linear physical size resolution of less that 2~pc for Local Group 
galaxies and of $\sim$30~pc at the Virgo cluster distance. 
Given the relatively small number and sparse distribution of luminous
X-ray sources expected in a normal galaxy (a few hundred at most in the 
Milky Way and M31, down to X-ray luminosities of $10^{36} \rm ergs/s$; 
Watson 1990; Supper at al 2001), 
confusion is seldom a problem when observing galaxies in the local Universe
with {\it Chandra}. Moreover, the small beam translates into very efficient
source detection, since the background contribution in the beam is 
virtually nil in most cases: 10 counts make a very significant source.
With these capabilities {\it Chandra} is opening up the field of X-ray population
studies in galaxies, and at the same time provides the best facility
for in-depth detailed studies of the diffuse hot interstellar emission.
These are the two aspects of {\it Chandra} results on galaxies that I will
discuss in this talk.

\section{X-ray Source Populations}

Population studies are often used in astronomy, to both constrain the
physical characteristics of a class of sources, and to study their 
evolution. While statistical studies of the global X-ray properties of
galaxies, and of their relation to the overall galaxian emission, 
could be pursued with pre-{\it Chandra} capabilities (e.g. see Fabbiano 1989, 
Eskridge et al 1995, Shapley et al 2001),
the study of the properties of the X-ray sources within a galaxy
had to be restricted to the most nearby galaxies.
Even so, it was clear that this approach had lots of potential.
As I first pointed out in 1995 (Fabbiano 1995), X-ray source population
studies are best done outside of the Milky Way: observations of
external galaxies provide complete samples of X-ray sources, without the
biases in distance and line of sight absorption inherent to Galactic
source studies. The first such studies with {\it Einstein}, followed 
by similar work with ROSAT, showed that the luminosity functions of X-ray 
sources may vary in different galaxies, although systematic effects, 
connected to morphology or star formation activity, could not be discerned 
at the time. A peculiar result, that was pointed out early on
(see Fabbiano 1989) and revisited periodically since, was the presence of ultra-luminous
or super-Eddington sources in some galaxies. These are sources
with X-ray luminosities well above the Eddington limit for an
accreting neutron star ($\sim 1.3 \times 10^{38} \rm ergs/s$), suggesting
the presence of 10-100s solar mass black holes. Although overall
one could not discriminate between single sources and clumps of emission,
in a few cases, variability and spectral variability results pointed
to compact X-ray binaries (Makishima 1994, Fabbiano 1995).

The first results with {\it Chandra} have indeed confirmed a variety 
of functional shapes for the X-ray Luminosity Functions (XLF) of
galaxian X-ray sources, and have also uncovered very significant
populations of Ultra-Luminous X-ray sources (ULX) in star-forming
galaxian regions. A brief summary of individual results follows.

\bigskip
\noindent
{\bf M33} -- The Local Group Sc galaxy M33 has been surveyed with ACIS 
on {\it Chandra} (McDowell et al. in preparation).
Since this galaxy covers an area of the sky significantly
larger than the ACIS field of view, a series of different pointings was
needed. So far $\sim 2/3$ of M33 has been observed 
with {\it Chandra}, resulting in $\sim 120$ sources, with luminosities
ranging from $\sim 10^{39} \rm ergs/s$ (the well-known luminous
X-ray nucleus nucleus), down to a
threshold luminosity of a few $\sim 10^{35} \rm ergs/s$.
Excluding the nucleus, all the point-like sources in this galaxy have 
sub-Eddington luminosities (for a 1~$M_{\odot}$ accretor).
Pending more accurate estimates of the completeness at lower
luminosities, there is a suggestion of flattening near the detection
threshold. Comparison of images in different spectral bands shows
a variety of X-ray colors (spectral parameters) for the X-ray sources.

\bigskip
\noindent
{\bf M81} -- The Sb galaxy M81, at a distance of $\sim$3.5~Mpc, has 
overall optical properties very similar to those of M31 (Andromeda),
from which it differs chiefly because of its active mini-Seyfert
nucleus, that is also the dominant X-ray source. {\it Einstein}
observations, however, suggested that M81 has an intrinsically more luminous
XLF than M31 (Fabbiano 1988a). {\it Chandra} ACIS observations of the
central 8.3'x8.3' field of this galaxy
(Tennant et al 2001) resulted in the detection of 97 point-like sources,
down to a limiting luminosity of $\sim 4 \times 10^{36} \rm ergs/s$.
Tennant et al derive XLFs for the X-ray sources
associated with the galactic disk and bulge separately.
These two XLFs differ: while the disk XLF follows a power-law
over the entire observed range, extending to super-Eddington luminosities,
the bulge XLF is steeper, flattening out at luminosities below
$\sim 10^{37} \rm ergs/s$. These different shapes confirm that
bulge and disk X-ray sources belong indeed to different populations,
as suggested from Galactic studies (see Watson 1990).

\bigskip
\noindent
{\bf E and SO galaxies} -- The presence of a bulge-like population of X-ray
binaries in early-type galaxies, and their effect on the overall
X-ray emission, was first suggested by Trinchieri \& Fabbiano (1985;
see also e.g. Fabbiano et al 1994, Pellegrini \& Fabbiano 1994, for
a discussion of their effect on the X-ray luminosity of X-ray 
faint E and S0s) and has been controversial till now.
Although deep ROSAT HRI observations detected some luminous
sources in NGC~5128 (Turner et al 1997) and NGC~1399 (Paolillo et al 2001),
it is only with {\it Chandra} that a clear, uncontroversial picture
of the X-ray emission of early type galaxies can be obtained.
A population of point-like sources is readily visible in the 
{\it Chandra} images of NGC~5128 (Centaurus~A), where 63 sources
were detected in the first HRC image (Kraft et al 2000). Rich populations of
point-like sources are also detected in the E NGC~4697 (Sarazin et al 
2001; Fig.~1)  and S0 NGC~1553 (Blanton et al 2001), 
and account for $\sim 1/2 - 2/3$ of the X-ray emission of these
X-ray-faint galaxies. The XLFs  of NGC~4697 and NGC~1553 both
have a break at $\sim 3 \times 10^{38} \rm ergs/s$, the Eddington
luminosity of a neutron star binary.
The {\it Chandra} image of NGC~1399 reveals a rich population of
luminous X-ray sources, prevalently associated with globular clusters
(Angelini et al 2001).

\begin{figure}
\centering
\plotfiddle{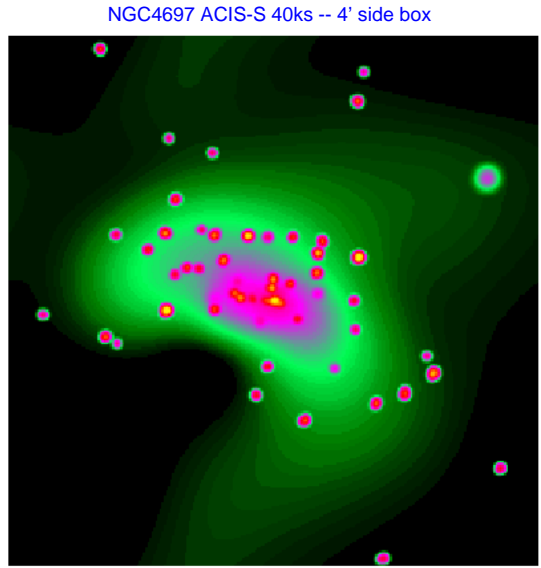}{8cm}{0}{50}{50}{-160}{-80}
\caption[]{{\it Chandra} ACIS image of NGC~4697, adaptively smoothed
with CIAO `csmooth'. Note the population of point-like sources and the
`warped' hot ISM.}
\end{figure}

\bigskip
\noindent
{\bf Actively Star-Forming Galaxies} -- The XLFs of starburst galaxies
tend to be skewed towards higher X-ray luminosities, when compared
with those of more aged systems. Examples include NGC~4038/9
(the Antennae; Fabbiano et al 2001; Zezas et al, in preparation), M82 (Zezas et al 2001),
NGC~3256 (Lira et al 2001) and
more, increasingly being added to the {\it Chandra} database.
Perhaps the most impressive XLF is that of the merging system
NGC~4038/9, where 49 point-like sources are detected with {\it Chandra}
with luminosities ranging from $10^{38}$ to $10^{40}$~ergs/s.
There are 14 point-like ULXs sources with $L_X > 10^{39}$~ergs/s.
A number of these sources are also detected in the nuclear region of M82,
including the exceptionally luminous, variable source, discussed
by Kaaret et al (2001) and Matsumoto et al (2001) and interpreted
as a `intermediate-mass' (100s $M_{\odot}$) black hole binary.
The X-ray spectra of the luminous sources in the Antennae are 
hard and can be fitted with a composite power-law + accretion disk
model, reminiscent of the spectra of 
ultra-luminous sources in more nearby galaxies
(e.g. Makishima 2000; Kubota 2001; La~Parola et al 2001).

Fig.~2 compares the XLFs of the
actively star-forming galaxies M82 and NGC4038/9 with the XLF of 
the elliptical galaxy NGC~4697 (Sarazin et al 2001) and that of the disk of M81
(Tennant et al 2001).
The XLF of NGC~4697 is definitely steeper than the others, and in this
resembles that of the bulge of M81 (Tennant et al 2001). This overall difference in shape
may be related to the lack of short-lived very luminous sources
in older stellar systems. What are these ultra-luminous sources?
Although variability and spectra suggest that they are compact
binary systems (Zezas et al in preparation), the jury is still out on the nature of the
accretor in these binaries: are they `intermediate-mass' black holes
(see above), or - more likely- less extreme objects, in beamed X-ray sources
(King et al 2001)?

\bigskip
I am sure that in the years to come the study of X-ray populations
in galaxies will become an increasingly used tool for understanding
properties and evolution of X-ray binaries and their relation to the
parent stellar population. Work is already underway on deep images
of spiral galaxies (e.g. the work by A. Prestwich and collaborators), 
and more data are sure to become available in the near future. 

%
%
%\begin{figure}
%\centering
%\plotfiddle{../COSPAR2000/cospar4.ps}{8cm}{-90}{50}{50}{-205}{295}
%\caption[]{Soft and Hard-band {\it Chandra} ACIS images of NGC~4038/9, adaptively smoothed
%with CIAO `csmooth'.}
%\end{figure}

%
\begin{figure}
\centering
\plotfiddle{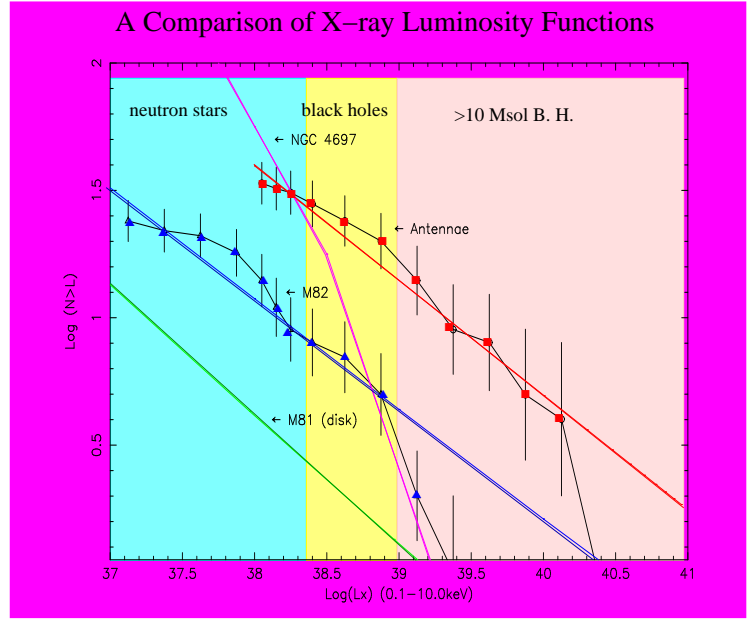}{6cm}{-90}{40}{40}{-140}{230}
\caption[]{X-ray source luminosity functions in different galaxies.}
\end{figure}

\section{Hot gaseous emission}

Hot Inter-Stellar Medium (ISM) can be uniquely detected and studied in X-rays.
This hot ISM is found in all kinds of galaxies, and its study may give us information
on (i) galaxy structure, e.g. the binding mass of Elliptical galaxies 
(e.g. Fabricant \& Gorenstein 1983); (ii) ecology, by measuring the flux of
metal-enriched material from galaxies to the intergalactic medium, via
galactic winds; and (iii) evolution, as a way for star formation to
affect the energy and entropy balance of clusters and their components 
(see simulations discussed in this meeting).
With {\it Chandra} we can separate the soft diffuse hot ISM emission from the 
harder point-source contribution both spatially and spectrally, and we can do detailed 
spatial/spectral studies of the hot ISM.

\subsection{Hot ISM and Superwinds in Starforming Galaxies}

Starburst galaxies can be spectacular X-ray emitters, since the
enhanced star formation activity results in enhanced X-ray
emission and in hot gaseous plumes and galaxian superwinds in the most extreme
cases (e.g. Fabbiano 1988b). Examples of this type of galaxies 
observed with {\it Chandra} include M82, NGC~4038/9 (Fabbiano et al 2001),
NGC~253 (Strickland et al 2000), and NGC~3256 (Lira et al 2001).

In NGC~4038/9,
besides the population of bright point-like sources already mentioned, we
detect a soft diffuse emission, with a typical optically thin plasma emission
spectrum (Fabbiano et al 2001) accounting for about 1/2 of the total X-ray 
emission ($\sim 1 \times 10^{41} \rm ergs/s$). This emission consists of
a patchy component associated with the star-forming knots,  and a lower surface brightness
component extending throughout the system. The extended emission component
can be detected farther South than the
stellar disks, suggesting a galaxian outflow or superwind. 

The angular resolution of the {\it Chandra} mirrors is such that it is
meaningful to pursue direct comparison between {\it Chandra} and
{\it Hubble} data. A comparison between the ISM component 
of NGC~4038/9 and the warm ISM visible in $H \alpha$ (using
archival HST WFPC data), suggests a complex and varied multi-phase ISM.
While there is a general resemblance of the spatial distribution of the X-ray 
and $H \alpha$ emission, we note both regions where the two surface 
brightnesses closely follow each others, and regions where hot superbubbles
fill-in holes in the $H \alpha$ distribution (Fabbiano et al 2001). 
The superbubbles we observe in the ISM of the Antennae galaxies
are extraordinary, if we compare them with similar features in the ISM
of more normal galaxies. Typical X-ray luminosities
of bright X-ray superbubbles are in the few$\times 10^{39} \rm ergs/s$
range, comparable with the entire thermal emission of the nucleus of the
nearby starburst galaxy NGC~253 (Fabbiano 1988b), and $\sim 20$ times
more luminous than 30~Dor.

%\begin{figure}
%\centering
%\plotfiddle{/data/pepi3/pepi/PAPERI_figs/fig9.ps}{7cm}{-90}{50}{50}{-185}{270}
%\caption[]{Left: {\it Chandra} X-ray image and contours; Right: {\it Chandra}
%contours over {\it HST WFPC} H$\alpha$ image. From Fabbiano et al (2001)}
%\end{figure}

These observations of nearby active star-forming galaxies provide
the detailed data needed to calibrate models of the effect of stellar 
formation and evolution on the ISM of galaxies. They also provide a 
local laboratory for the phenomena occurring at the epoch of galaxy formation.

\subsection{The `Feature-rich' ISM of Early-type Galaxies}

The presence of large amounts of ISM in E and S0 galaxies was an
X-ray discovery. This ISM is hot and was first convincingly 
detected with the {\it Einstein Observatory}. 
A short history of the observations and modelling of this ISM, and 
the ensuing controversies can be found in Fabbiano \& Kessler (2001).
With {\it Chandra}, we are obtaining a new detailed look that reveals
a wealth of structure in what were by-and-large considered
fairly uniform hot halos shaped by the galaxy's gravitational
field. Examples of `disturbed' hot halos are provided by the
{\it Chandra} observations of NGC~4636, NGC~5044 and NGC~4697.

\medskip
\noindent
{\bf NGC~4636} -- This Virgo elliptical is one of the galaxies
with largest, most extended X-ray halos. The {\it Chandra} observation
(proposed by R. Mushotzky; data are now in the public {\it Chandra} archive),
reveals both a rich population of individual galactic X-ray sources,
including a ring-like concentration in the nuclear region, and a strongly defined,
8 kpc long, spiral-like feature in the X-ray halo (Fig.~3). Recently, Jones et al
(2001) propose that this structure may be due to shocks
driven by a past nuclear outburst. Ciotti and Ostriker (1997, 2001)
developed a model in which recurrent outburst and cooling flow
accretion episodes may occur in early-type galaxies, giving 
rise to AGN/quiescent cycles.

\medskip
\noindent
{\bf NGC~5044} -- This elliptical galaxy, observed by P. Goudfrooij, and also in the {\it Chandra}
public archive, provides another example of a large, centrally disturbed, X-ray halo (Fig.~4).
In this case, a central radio source is present, and the X-ray
feature could be related to it. A disturbed inner halo is also present in
NGC~1399, another early-type galaxy with nuclear radio emission, where some
of the features suggest interaction between radio lobes and hot ISM
(Paolillo et al 2000).

\begin{figure}
\centering
\plotfiddle{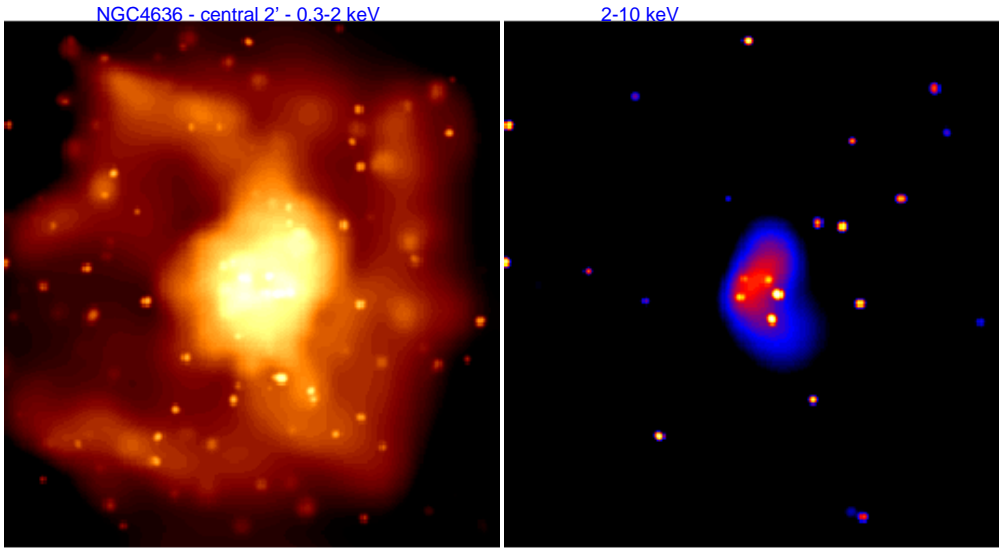}{7cm}{0}{70}{70}{-200}{-180}
\caption[]{Left: Soft (0.3-2~keV) {\it Chandra} ACIS image of the central regions of
NGC~4636, showing the hot halo and the spiral feature of the hot ISM. Right: Image in the
2-10~keV band, showing the hard point source (also visible in the soft image).}
\end{figure}

\begin{figure}
\centering
\plotfiddle{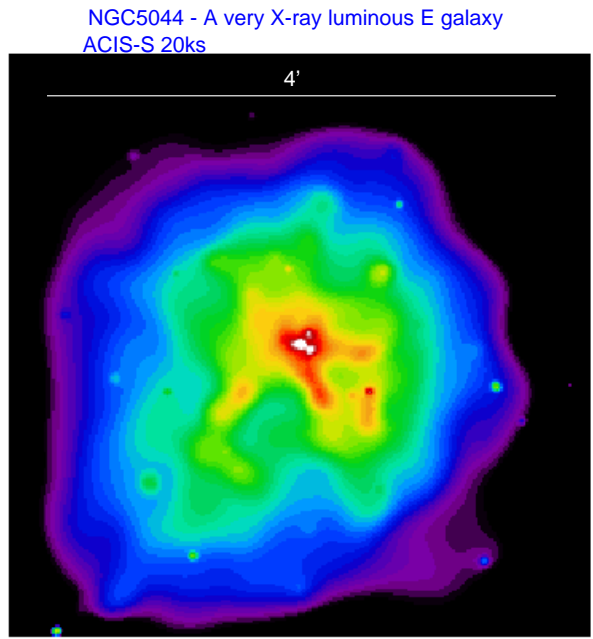}{8cm}{0}{60}{60}{-200}{-113}
\caption[]{{\it Chandra} ACIS image of NGC~5044. Data were adaptively smoothed with
`csmooth'.}
\end{figure}

\medskip
\noindent
{\bf NCG~4697} -- This is the X-ray faint galaxy whose X-ray source population
was discussed earlier in this paper (Fig.~1). As reported by Sarazin et al (2001), the {\it Chandra}
image also shows a soft extended component of the X-ray emission.
This hot ISM does not follow the general symmetry of the stellar distribution,
but it is both more extended in its radial behaviour and warped, suggesting some
dynamical interaction.

\medskip
Clearly these are only a few first results. It is not difficult to predict,
based on these data, that {\it Chandra} observations of the gaseous component of
E and S0 galaxies will provide novel insight on their evolution and physical 
properties.

\section{Conclusions}

It is certainly way too soon to evaluate the effect of high spatial/spectral
resolution {\it Chandra} X-ray observations in our understanding of galaxies and their
environment, since only a very small amount of data have been fully analyzed and
more observations are continually being done. However, a first look shows 
that {\it Chandra} has more than fulfilled its promises.

High resolution spectrally-resolved images have solved many outstanding issues,
in a very direct and simple way. Most of these issues were connected with the
interpretation of lower resolution data of E and S0 galaxies. I am pleased to 
say that the conclusion, put forward by me and my collaborators, that all
E and S0 galaxies have a baseline X-ray emission from point-like X-ray binaries
in addition to a varying amount of hot ISM
(e.g. Trinchieri and Fabbiano 1985;  Eskridge et al 1995; 
Pellegrini \& Fabbiano 1994) has withstood the high resolution test and is 
no longer controversial.
{\it Chandra} images are also revealing interesting features in the hot halos of these
galaxies, that suggest either external interactions or the effect of nuclear 
activity. Studies of the X-ray populations of galaxies, that can provide a
direct probe of the massive stellar component and its evolution, as well as
of the properties of matter in its most extreme form (neutron stars, black holes),
were merely an intriguing future possibility (see Fabbiano 1995) and are
now reality. Given the relatively sparse nature of luminous X-ray sources
in galaxies, individual sources can be now studied in galaxies as far as Virgo
and beyond, where only population synthesis can be performed of the normal
stellar population.

With {\it Chandra}, X-ray observations have become an important tool of main-stream
astrophysical research.

\acknowledgments
This work was supported by NASA contracts NAS8-39073 (CXC) and
NAS8-38248 (HRC).

\end{document}